\documentclass[12pt]{article}

\def\la{\mathrel{\mathpalette\fun <}}
\def\ga{\mathrel{\mathpalette\fun >}}
\def\fun#1#2{\lower3.6pt\vbox{\baselineskip0pt\lineskip.9pt
\ialign{$\mathsurround=0pt#1\hfil##\hfil$\crcr#2\crcr\sim\crcr}}}

\begin{document}
\renewcommand{\theequation}{\thesection.\arabic{equation}}

\begin{titlepage}
\renewcommand{\thefootnote}{\fnsymbol{footnote}}

\begin{flushright}
PNPI-TH-3400\\
hep-th/0005088\\

\end{flushright}

\vfil

\begin{center}
\baselineskip20pt
{\bf \Large  What Do We Learn about Confinement from \\
the Seiberg-Witten Theory \footnote{Talk given at XXXIV PNPI
Winter School in Nuclear and Particle Physics,
 Repino, St.Petersburg, Russia, February 14--20, 2000}}
\end{center}
\vfil

\begin{center}
{\large   Alexei Yung}

\vspace{0.3cm}

{\it Petersburg Nuclear Physics Institute, Gatchina, St. Petrsburg
188350}

\vfil

{\large\bf Abstract} \vspace*{.25cm}
\end{center}

The confinement scenario in $N=2$ supersymmetric gauge theory at
the monopole point is reviewed. Basic features of this
$U(1)$ confinement are contrasted with those we expect in QCD.
In particular, extra states in the hadron spectrum and
non-linear Regge trajectories are discussed. Then another
confinement scenario arising on Higgs branches of the theory
with fundamental matter is also reviewed.
Peculiar properties of the Abrikosov--Nielsen--Olesen string on
the Higgs branch lead to a new confining regime with the
logarithmic suppression of the linear rising potential.
Motivations for a search for tensionless strings are proposed.

\vfil

%\begin{flushleft}
%April 2000
%\end{flushleft}
\end{titlepage}

\newpage

\section{Introduction}

Ideas of electromagnetic  duality \cite{1} lead to a dramatic
breakthrough in our understanding of the dynamics of strongly
coupled supersymmetric gauge theories. Particularly spectacular
results were obtained by Seiberg and Witten in ${\cal N}=2$
supersymmetry where the low energy effective Lagrangians were
found exactly \cite{SW1,SW2}.

One of the most important physical outcomes of the
Seiberg--Witten theory is the demonstration of  confinement
of charges via the monopole condensation. The scenario for
confinement as a dual Meissner effect was proposed by Mandelstam
and 't Hooft many years ago \cite{MH}. However, because the
dynamics of monopoles is hard to control in non-supersymmetric
gauge theories this picture of confinement remained as an
unjustified qualitative scheme.

The breakthrough in this direction was made by Seiberg and
Witten \cite{SW1,SW2}. Using the holomorphy imposed by ${\cal N}=2$
supersymmetry they showed that the condensation of monopoles to
really occurs near the monopole point on the modular space of
the theory once ${\cal N}=2$ supersymmetry is broken down to ${\cal N}=1$ one
by the mass term for the adjoint matter.

In this talk I will review the confinement scenario in the
Seiberg--Witten theory underlining the basic features of this
$U(1)$ confinement which distinguish it from those we expect in
QCD.

In particular, I am going to discuss extra hadron states arising
in the Seiberg--Witten theory which we do not expect in QCD or
in ${\cal N}=1$ supersymmetric QCD.

Then I focus on Abrikosov--Nielsen--Olesen (ANO) strings
\cite{ANO} which are responsible for the confinement near the
monopole point. We will see that these strings turn out to be
too "thick" and cannot be described by the standard string theory
approximation of long and infinitely thin strings. The result of
this is that Regge trajectories become linear only at very large
values of spin $j$.

In the second part of my talk I will review another confinement
scenario arising in the Seiberg--Witten theory with the fundamental
matter: confinement on Higgs branches \cite{SW2}. Higgs branch
represents a limiting case of the superconductor of type I with
vanishing Higgs mass. We will see that in this limit ANO
vortices becomes logarithmically "thick" \cite{Y}. Because of
this the confining potential is not linear any longer. It
behaves as $L/\log L$ with the distance between heavy trial
charges (monopoles). This peculiar confining regime can occur
only in supersymmetric theories.

In the end of my talk I will  speculate on the possible ways to
avoid at least some of the unwanted features of
$U(1)$ confinement I am discussing in this talk.

\section{Confinement as a dual Meissner effect}
\setcounter{equation}{0}

First let me remind  the mechanism of confinement suggested by
Mandelstam and 't Hooft \cite{MH}. Consider an Abelian--Higgs
model with the action
\begin{equation}
S_{AH}=\int d^4x\left\{\frac1{4g^2}\,F^2_{\mu\nu}+|\nabla_\mu
\varphi|^2+\lambda(|\varphi|^2-v^2)^2\right\}.
\end{equation}
Here $\varphi$ is a complex scalar field,
$\nabla_\mu=\partial_\mu-in_eA_\mu$, where $n_e$ is the electric
charge of $\varphi$. We assume the weak coupling $g^2\ll1$. The
scalar field in (2.1) develop VEV
\begin{equation}
|\langle\varphi\rangle |\ =\ v\ ,
\end{equation}
which breaks $U(1)$ gauge group. The photon acquires the mass
\begin{equation}
m^2_\gamma\ =\ 2n^2_eg^2v^2\ ,
\end{equation}
while the mass of the Higgs boson (one real degree of freedom)
is
\begin{equation}
m^2_H\ =\ 4\lambda v^2\ .
\end{equation}

Now introduce an infinitely heavy trial monopole and
anti-monopole in the vacuum of the Abelian Higgs model. We can
think of them as of Dirac monopoles or as of 't Hooft--Polyakov
monopoles of some underlying non-Abelian gauge theory broken
down to $U(1)$.

Monopoles has quantized values of magnetic flux $2\pi n/n_e$.
This magnetic flux cannot  be absorbed by the vacuum once there
are no dynamical magnetic charges in the theory.
 On the other hand, magnetic field cannot penetrate into the Higgs
vacuum. As a result ANO vortex appears \cite{ANO} connecting
monopole with anti-monopole. This vortex can be viewed as a 
bubble of an unbroken vacuum inside the Higgs vacuum.

It has $\varphi=0$ along the line connecting monopole and
anti-monopole and magnetic flux $2\pi n/n_e$, $n$ --- winding
number. ANO vortex is a solution of equations of motion for the
Abelian--Higgs model (2.1). It corresponds to a non-trivial map
from the infinite circle in the plane orthogonal to the axis of
the vortex to the gauge group, $\pi_1(U(1))=Z$. For example, for
the winding number $n=1$ fields $\varphi$ and $A_\mu$ behave at
the infinity as
\begin{eqnarray}
&& \varphi\ \sim v e^{i\alpha}\ , \nonumber\\
&& A_m\ \sim\ -\varepsilon_{mn}\frac{x_n}{x^2}\ ,
\end{eqnarray}
where $x_n$, $n=1,2$ is the distance from the vortex axis in the
plane orthogonal to this axis, while $\alpha$ is the polar angle
in this plane.

Because the vortex has a fixed energy per unit length (string
tension $T$) the potential between monopole and anti-monopole at
large distances $L$ behaves as
\begin{equation}
V(L)\ =\ TL\ .
\end{equation}
This linear potential means confinement of monopoles.

The ratio of the Higgs mass (2.4) to the photon mass (2.3) is an
important parameter characterizing the type of superconductor.
If $m_H>m_\gamma$ we have the type II superconductor. Different
vortices interact via repulsive forces. In particular, the
tension of the vortex with winding number $n$ is larger than the
sum of tensions of $n$ vortices with winding numbers $n=1$.
Therefore, vortices with higher winding numbers, $n>1$ are
unstable.

In particular, for $m_H\gg m_\gamma$ (London limit)
 the vortex solution can be
found in the analytic form. The string tension for this case was
calculated by Abrikosov in 1957 and later on re-obtained by
Nielsen and Olesen in 1973 \cite{ANO} in the framework of the
relativistic field theory. The result for $n=1$ is
\begin{equation}
T\ =\ 2\pi v^2\ln\frac{m_H}{m_\gamma}\ .
\end{equation}

For the particular interesting case $m_H=m_\gamma$ vortices
satisfy the first order equations. They saturate the Bogomolny bound
and their string tension reads \cite{B}
\begin{equation}
T_n\ =\ 2\pi v^2n\ ,
\end{equation}
where $n$ is the winding number.
In particularly, as is clear from (2.8), different vortices do
not interact.

In supersymmetric theories the BPS-saturation means that some of
SUSY generators act trivially on the vortex solution \cite{HS}.
The Bogomolny bound (2.8) coincides with the central charge of
SUSY algebra. This means that the classical result (2.8) for
 the vortex
string tension remains exact in the quantum theory.

For $m_H<m_\gamma$ we have type I superconductor. In this case
vortices attract each other. In particular, vortices with
higher $n$ are stable \cite{BV}. For the case $m_H\ll
m_\gamma$ the vortex solution can be found analytically
\cite{Y}. The string tension in this case has the form \cite{Y}
\begin{equation}
T_n\ =\ \frac{2\pi v^2}{\ln m_\gamma/m_H}\ ,
\end{equation}
and does not depend on $n$ to the leading order in $\ln
m_\gamma/m_H$.

In the general case the string tension $T_n$ is a monotonic 
function of the ratio $m_{H}/m_{\gamma}$ \cite{BV}. It reaches
its Bogomolny bound (2.8) at $m_H=m_\gamma$.

To conclude this section let me point out that the main lesson
to learn here is
 that the condensation of electric charges cause the
confinement of monopoles. Vice versa,  the condensation of
monopoles in the dual Abelian Higgs model leads to the
confinement of electric charges.

\section{Monopole condensation}
\setcounter{equation}{0}

Now consider ${\cal N}=2$ gauge theory. Most of all in this talk I will
be talking about the simplest case of the theory with
$SU(2)$ gauge group
studied in \cite{SW1}. The simplest version of this theory contains 
one 
  ${\cal N}=2$ vector multiplet.
This multiplet on the component level consists of  the
gauge field $A^a_\mu$, two Weyl fermions $\lambda^{\alpha a}_1$
and $\lambda^{\alpha a}_2$ $(\alpha=1,2)$ and the complex scalar
$\varphi^a$, where $a=1,2,3$ is the color index.

The scalar potential of this theory has a flat direction. Thus
the scalar field can develop an arbitrary VEV along this
direction breaking $SU(2)$ gauge group down to $U(1)$. We choose
$\langle\varphi^a\rangle=\delta^{a3}\langle a\rangle$. The
complex parameter $\langle a\rangle$ parameterize the moduli
space of the theory (Coulomb branch). The low energy effective
theory contains only the photon $A_\mu=A^3_\mu$ and its
superpartners: two Weyl fermions $\lambda^3_1$, $\lambda^3_2$
and the complex scalar $a$.

The Coulomb branch can be parameterized by the gauge invariant parameter
$u=\langle\varphi^{a2}/2\rangle$. It
 has two singular points $u=\pm2\Lambda^2$
( $\Lambda$ is the scale of
gauge theory \footnote{We use the Pauli-Villars regularization scheme
here.}) where monopole or dyon becomes massless. Near,
say, the monopole point $(u=2\Lambda^2)$ the effective low
energy theory is  dual ${\cal N}=2$ QED. This means that the theory
has light monopole hypermultiplet interacting with the dual
photon multiplet in the same way as ordinary charges interact
with the ordinary photon. The action of this dual QED reads
\begin{equation}
S_{eff}=\ S^{eff}_g+S^{eff}_m\ .
\end{equation}
Here the action for the gauge field is
\begin{equation}
S^{eff}_g=\int d^4x\left\{\int d^2\theta d^2\bar\theta \frac1{g^2}
\bar A_DA_D+\int d^2\theta\frac1{4g^2}W^2_D+c.c.\right\},
\end{equation}
where $A_D$ is the  dual chiral ${\cal N}=1$ field. Its lowest component
$a_D$ goes to zero at the monopole point. $W_D^\alpha$
$(\alpha=1,2$) is the ${\cal N}=1$ chiral field of the dual photon
field strength.  Together $A_D$ and $W_D$ form  ${\cal N}=2$ vector
U(1) multiplet.

The matter-dependent part of the action reads
\begin{eqnarray}
&& S^{eff}_m=\int d^4xd^2\theta d^2\bar\theta\left[\bar Me^{V}M
+\bar{\tilde M}e^{-V}\tilde M\right]\ + \nonumber\\
+ && i\int d^4xd^2\theta\sqrt2\tilde M A_DM\ +\ c.c.\ .
\end{eqnarray}
Here $M,\tilde M$ are two chiral fields of the monopole hypermultiplet.
The monopole mass (given by $m^2_m=2|a_D|^2$) goes to zero at the
monopole point $a_D=0$.

Now let us break  ${\cal N}=2$ QED (3.1) down to ${\cal N}=1$ adding
 the mass term for the
adjoint matter in the microscopic SU(2) Seiberg--Witten theory 
\begin{equation}
S_{\rm mass}\ =\ i\int d^4xd^2\theta\,\mu \Phi^{a2}+\ c.c.\,
\end{equation}
where $\Phi^a$ is the ${\cal N}=1$ chiral superfield which contains
component fields $\varphi^a$ and $\lambda^{\alpha a}_2$.
 Expressed in
terms of $A_D$  near the monopole point in the effective
theory it reads \cite{SW1}
\begin{equation}
\mu \Phi^{a2}=\ -\sqrt2\,\xi A_D+\frac{\mu_{D}}{2} A^2_D+O(A^3_D)\ ,
\end{equation}
where
\begin{eqnarray}
\xi & = & 2i\; \mu \Lambda\ , \nonumber\\
\mu_{D} & = &  -\frac{27}{4}\; \mu \ .
\end{eqnarray}
The coefficients in (3.6) can be read off the Seiberg--Witten
exact solution \cite{SW1}. 

Minimizing the superpotential in (3.1),(3.5) with respect to
$M,\tilde M$ and $A_D$ we find that the Coulomb branch shrinks
to the point
$$
\langle a_D\rangle\ =\ 0\ ,
$$
while
\begin{equation}
\langle \tilde mm\rangle\ =\ \xi\ ,
\end{equation}
where $m,\tilde m$ are the scalar components of $M,\tilde M$.
Taking into account  the $D$-term condition in (3.1)
\begin{equation}
|\langle m\rangle|\ =\ |\langle\tilde m\rangle|
\end{equation}
we get the monopole condensate
\begin{equation}
|\langle m\rangle|\ =\ |\langle\tilde m\rangle|\ =\ \sqrt{|\xi|}\ .
\end{equation}
The monopole condensation breaks
 the U(1) gauge group and  ensures 
confinement of electric charges. ANO strings arising as a result
of  the monopole condensation connect quarks with anti-quarks (we
interpret these states as mesons) or with another quarks with
opposite electric charge (we interpret these states as
baryons).

Note, that the effective  dual QED is in the weak coupling regime at small
$\mu $. The QED coupling behaves as
\begin{equation}
\frac{8\pi^2}{g^2}\ \sim\ -\log\frac{\mu }\Lambda\ .
\end{equation}
Moreover at $\mu\ll\Lambda$ we can ignore non-Abelian effects. Note,
that $W$-boson mass $m^2_W=2|a|^2$ is of order of $\Lambda$ at
the monopole point, $m_W\sim\Lambda$.

Parameters $\xi$ and $\mu_{D}$ in the mass term perturbation
(3.5) play quite
different role in the effective QED description of the theory.
As it is noted in \cite{HSZ} the linear term in (3.5)
is the Fayet--Iliopoulos $F$-term which do not break ${\cal N}=2$
supersymmetry.  I will explain this in more details in the next
section.

On the contrary, the mass term for $A_D$ in (3.5) proportional to $\mu_{D}$
breaks ${\cal N}=2$ supersymmetry because it shifts the mass of $A_D$
away from the photon mass.

\section{The U(1) confinement versus the QCD--like
confinement}
\setcounter{equation}{0}

Now let me discuss several basic features of the confinement near
the monopole point in the Seiberg--Witten theory 
at small $\mu$ and contrast
them to those we expect in QCD. It is believed that non-supersymmetric
Yang-Mills theory  is in the same universality class as ${\cal N}=1$
Yang-Mills theory. The latter can be obtained as 
a large $\mu $ limit of the theory under consideration.
The reason is that in this limit the adjoint matter
 becomes heavy and decouples.
 Therefore, we  also expect the QCD-like
confinement at large $\mu$ in the theory at hand.
Unfortunately we have no control over the theory in this limit.

 Now I will
show that the $U(1)$ confinement in the theory at small $\mu$
has several important distinctions from what we expect from the theory
at large $\mu$. 

\subsection{Higher winding numbers}

The first problem I would like to talk about arises already in
SU(2) theory \cite{S}. As we discussed before the
flux of the ANO vortex is given by its winding number $n$, which
is an element of $\pi_1(U(1))=Z$ (for SU$(N_c)$ group it is
$\pi_1(U^{N_c-1}(1))=Z^{N_c-1})$. This could produce an extra
multiplicity in the hadron spectrum which we do not expect in
QCD. In QCD or in the large $\mu$ limit of the present theory we
expect classification of states under the center of the gauge
group, $Z_2$ for SU(2) rather than $Z$ ($Z_{N_c}$ for
SU$(N_c)$).

Consider as an example  ANO vortex with $n=2$
in the effective dual QED at small $\mu$. This
 string connects two quarks with two anti-quarks producing
an "exotic" state.  Note, that the string with $n=2$, in principle,
can be broken by $W$-boson pair creation but at low energies
 we
can neglect this process because $W$-boson is too heavy
at small $\mu$, $\mu \ll\Lambda$ ,
($m_W\sim\Lambda$).

The discussion above of "exotic" states in the hadron spectrum
\cite{S} is based on the purely topological reasoning. Now let
us consider the dynamical side of the problem. As we have seen
in Sect.2, strings with higher $n$ are stable or unstable
depending on the type of the superconductor. Namely,  in type I
superconductor the energy of the vortex with winding number $n$
is less than the total energy of $n$ vortices  with winding
numbers $n=1$.  Therefore vortices  with $n>1$ are stable and we
really have an "exotic" states in the spectrum.

On the contrary, in type II superconductor vortices with $n>1$
are unstable against decay into $n$ vortices with winding
numbers $n=1$. Therefore, "exotic" states are unstable and
actually in the ``real world'' at strong coupling might be not
observable at all.

Thus the natural question arises: what is the type of
superconductivity at the monopole point in the Seiberg--Witten
theory. This problem is studied in \cite{VY}. Let me briefly
present the result.

To the leading order in $\mu /\Lambda$ we ignore the mass
term for $A_D$ in (3.5) and consider only the linear in $A_D$
term parameterized by $\xi$. As I mentioned before this term is
the generalized Fayet--Illiopoulos (FI) term.
Let me explain what does this mean.
Let us start with  ${\cal N}=1$ QED.
 In  ${\cal N}=1$ 
supersymmetric U(1) gauge  theory one can add FI term to the action
\cite{FI} (we call it FI $D$-term here)
\begin{equation}
-\xi_3\ D\ ,
\end{equation}
where $D$ is the $D$-component of the gauge field. In ${\cal N}=2$ SUSY
theory field $D$ belongs to the $SU_R(2)$ triplet together with
$F$-components of the field $A_D$, $F_D$ and $\bar F_D$. Namely,
let us introduce the triplet $F_a$ $(a=1,2,3)$ using relations
\begin{eqnarray}
D &=& F_3\ , \nonumber\\
F_D &=& \frac1{\sqrt2}\ (F_1+iF_2)] , \nonumber\\
\bar F_D &=& \frac1{\sqrt2}\ (F_1-iF_2)\ .
\end{eqnarray}
Now the generalized FI-term can be written as
\begin{equation}
S_{FI}\ =\ -i\int d^4x\ \xi_aF_a\ .
\end{equation}
Comparing this with (3.5) we identity
\begin{eqnarray}
\xi &=& \frac12\ (\xi_1-i\xi_2)\ , \nonumber\\
\bar\xi &=& \frac12\ (\xi_1+i\xi_2)\ .
\end{eqnarray}
For this reason  we call the term linear in $A_D$ in (3.5)
FI $F$-term.

It is well known that the ${\cal N}=1$ QED  with FI $D$-term has BPS
ANO string \cite{HS,DDT,GS}. The reason for this is the following.
After the breakdown of $U(1)$ gauge group the photon acquires
the mass given by
(2.3) $(v^2=2|\xi|$ in (2.3), see (3.9)). The massive
vector ${\cal N}=1$ supermultiplet contains, in particular, one real
scalar. This scalar plays the role of the Higgs boson in the
Abelian Higgs model (2.1). Thus the BPS condition $m_H=m_\gamma$
is imposed by supersymmetry.

Now return to ${\cal N}=2$ supersymmetry and consider dual
QED (3.1) with the FI $F$-term added.
 The FI parameter $\xi^a$
explicitly breaks the $SU_R(2)$ group. However, the ${\cal N} =2$
supersymmetry remains unbroken. To see this note that FI term
is proportional to   $F$ and $D$ components of the vector multiplet
which transform as a total derivatives under the ${\cal N} =2$
supersymmetry transformation.

It is clear that  the FI $F$-term which appears in the
Seiberg--Witten theory (see (3.5)) can be obtained by $SU_R(2)$
rotation from the FI $D$-term. Moreover, masses of particles in
${\cal N} =2$ multiplet do not change under this rotation.
In particular, the condition
$m_H=m_\gamma$ stays intact. Hence, the ANO
string is BPS-saturated   \cite{HSZ,Sp,VY}.
Its string tension is given by (2.8). For $n=1$
\begin{equation}
T\ =\ 2\pi(2|\xi|)\ ,
\end{equation}
where we use that $v^2=2|\xi |$.

To consider the next-to-leading correction in $\mu /\Lambda$
we switch on   the mass
term for $A_D$  parameterized by $\mu_{D}$
in the effective theory, see (3.5)
. It breaks the ${\cal N}=2$ supersymmetry and
split the ${\cal N}=2$ multiplet. In particular, it breaks the
BPS condition $m_H=m_\gamma$. The effect of this term is studied
in \cite{VY}. The result is that the Higgs mass in the effective
Abelian Higgs model appears to be less than the photon mass,
$m_H<m_\gamma$ and the theory is driven to the type I
superconductivity. The string tension is less than its Bogomolny
bound,
\begin{equation}
T\ <\ 2\pi(2|\xi|)\ .
\end{equation}
In particular, in the large $\mu_{D}$-limit $\mu_{D}^2\gg\xi$ the string
tension is found analytically \footnote{To  take this limit
 we ignore relations (3.6), (3.10) and 
consider  QED (3.1) with the prturbation (3.5)
on its own right viewing parameters  $\xi$ , $\mu_{D}$ and $g^2$
as independent ons, assuming only the weak coupling condition
$g^2\ll 1$\cite{VY}.}\cite{VY}:
\begin{equation}
T\ =\ \frac{2\pi(2|\xi|)}{\ln(\frac{g |\mu_{D}|}{2\sqrt{|\xi|}})}\ .
\end{equation}

As I explained above
the result in (4.6) means that we really have an infinite tower of
"exotic" hadron states with higher string fluxes near the
monopole point at small $\mu $.

Of course, if we increase $\mu $ and approach
$\mu \sim\Lambda$  the unwanted strings with $n>1$ become
broken by the $W$-bosons production. The only string with $n=1$
connecting one quark with one anti-quark\footnote{Or with another
quark with the opposite electric charge.} will probably survive.
This string is believed to be responsible for the confinement in
${\cal N}=1$ SQCD at large $\mu $ .
It is called QCD string. This picture of confinement is proposed
in refs.\cite{W,HSZ}
 within the brane approach.

However, this scenario is hard to implement in the field
theory. One reason for this is that at large $\mu $ dual QED
enters the strong coupling regime and is no longer under control.
 Another one is probably even
more fundamental. The point is that  the role of matter fields
in the effective QED (3.1) is played by monopoles. As $\mu $
approaches $\Lambda$ the inverse mass of the dual photon (2.3)
approaches the size of monopole (which is of order of the inverse
$W$-boson mass $m^{-1}_W\sim\Lambda^{-1}$). Under these
conditions we hardly can consider monopoles as a local degrees
of freedom and the dual QED effective description breaks down.
In particular, we don't have a field theoretical description of
$n=1$ string in the region of large $\mu \ge\Lambda$.

\subsection{p--strings}

Another problem of $U(1)$ confinement in Seiberg--Witten theory
was noticed by Douglas and Shenker \cite{DS}. It appears in
$SU(N_c)$ gauge theories at $N_c\ge3$.

Scalar VEV's breaks the gauge group down to $U(1)^{N_c-1}$.
Hence, $N_c-1$ different types of ANO vortices arise  each
one associated with a particular
$U(1)$ factor. Their string tensions are given by \cite{DS}
\begin{equation}
T_p\ \sim\ \xi\sin\frac{\pi p}{N_c}\ ,
\end{equation}
where $p=1,\ldots, N_c-1$ numerates different $U(1)$ factors..
They are called $p$-strings. The number of strings with
different string tensions equals to $[(N_c-1)/2]$.

Therefore it is clear that extra states emerge in the hadron
spectrum which we do not expect in QCD \cite{DS}. Namely, the
number of quark--anti-quark meson states is $N_c$. Let me explain
this for the case of $SU(3)$.

We take three different quarks of $SU(3)$ as
\begin{equation}
q_1=\left(\begin{array}{l}1\\0\\0\end{array}\right),  \quad
q_2=\left(\begin{array}{l}0\\1\\0\end{array}\right), \quad
q_3=\left(\begin{array}{l}0\\0\\1\end{array}\right)
\end{equation}
and choose generators of two $U(1)$ groups of the broken
$SU(3)$ to be
\begin{equation}
\left(\begin{array}{ccc}1&0&0\\ 0&-1 &0\\ 0&0&0
\end{array}\right)
\quad \mbox{ and }\quad
\left(\begin{array}{ccc}0&0&0\\ 0&1 &0\\ 0&0&-1
\end{array}\right) .
\end{equation}
Thus, three quarks in (4.9) have the following electric charges
with respect to two $U(1)$ groups:
\begin{equation}
(1,0)\ ; \quad (-1,1)\ ; \quad (0, -1)\ .
\end{equation}
Now it is clear that $\tilde q_1q_1$ meson is formed by 1-string,
$\tilde q_2q_2$ meson is formed by 2-string and 1-anti-string and
$\tilde q_3q_3$ mesons is formed by 2-anti-string. In sum we have 3
meson states. Two of them ($\tilde q_1q_1$ and $\tilde q_3q_3$) are
(classically) degenerative while the third one, $\tilde q_2q_2$ is
two times heavier than the first two \footnote{This is true if
we neglect masses of quarks at the ends of strings (see
subsection 4.4 for the discussion of this point).}

In general we have $[(N_c+1)/2]$ families of $\tilde qq$ mesons
with different masses \cite{DS,HSZ}. Each family contains 
two mesons (one family contains one meson if $N_c$ is odd).
 They are classically degenerative but can split in
quantum theory.

Let me illustrate this splitting for the simplest example of
$SU(2)$ gauge group. For $SU(2)$ we have two classically degenerative
mesons. If we decompose the
quark as
\begin{equation}
q=q_+\left({1 \atop 0}\right)+q_-\left({0 \atop 1}\right),
\end{equation}
then these two mesons are $\tilde q_+q_+$ and $\tilde q_-q_-$.
Out of these states  we
can form two different combinations as follows
\begin{eqnarray}
&& \tilde qq\ =\ \tilde q_+q_+ +\tilde q_-q_-\ , \nonumber\\
&& \tilde q\tau_3q\ =\ \tilde q_+q_+ -\tilde q_-q_-\ .
\end{eqnarray}
Here $\tau_3$ is the color matrix. In gauge invariant notation
two states in (4.13) are
\begin{equation}
\tilde qq\ , \quad \frac1{\sqrt{\varphi^2}}\ \tilde
q\varphi q\ .
\end{equation}
It is clear that
classically these states are degenerative but in quantum theory
they split. The heavier state acquires a large width and might
not be observable at all.

Still we have $[(N_c+1)/2]$ different $\tilde qq$ meson states
instead of one state we expect in the large $\mu $ limit.

As it is noted in \cite{DS} $W$-bosons are charged under
different $U(1)$ factors and therefore provide a coupling between
them. Thus, we expect that at large $\mu $, $\mu \sim
\Lambda$ most of $\tilde qq$-mesons discussed above
 become unstable and disappear.
In fact, it is shown in \cite{HSZ} within the brane approach
that as we increase $\mu $ all $\tilde qq$-meson disappear
except one in which quark is connected with anti-quark by the
1-string. This string is shown \cite{HSZ} to become QCD string
of ref.\cite{W} at large $\mu $.

Unfortunately, as I mentioned before, we still have no
description of this transition in  field theory.

$P$-strings by themselves also survive the large $\mu $-
limit \cite{HSZ}. They connect $p$ quarks in the $p$-index
antisymmetric representation with $p$-anti-quarks to form an
"exotic" meson. Also they can form  baryons.

To conclude this subsection I would like to mention the recent
paper \cite{Sp2} in which ${\cal N}=2$ supersymmetry breaking
with the first two Casimir operators is considered
for the $SU(N_{c})$ theory. It is claimed that
once off-diagonal couplings between different $U(1)$ factors
are taken into account $p$-strings in generic case 
fail to be BPS saturated
even in the limit of zero $\mu_{D}$.

\subsection{Non-linear Regge trajectories}

One of the main motivations to consider hadrons as quarks
connected by strings in early days of  String Theory was
the linear behavior of Regge trajectories. However, as I show
now the ANO string at the monopole point of the  Seiberg--Witten
theory does not produce linear Regge trajectories. More
precisely Regge trajectories become linear only at very large
$j$.

Let me first show the linear behavior of Regge trajectories
using very elementary classical arguments.

Consider long ANO string (of the length $L$) connecting light
quark and anti-quark \footnote{See next subsection for the
discussion on whether quarks can be light near the monopole
point.}. Let this string to rotate around the axis
orthogonal to the string with large spin $j$. If $j$ is large
enough the problem becomes classical and we can apply classical
equations of motion to the rotating string. For the linear
potential
\begin{equation}
V(L)\ =\ TL
\end{equation}
equation of motion looks like
\begin{equation}
T\ \sim\ E\omega^2L\ ,
\end{equation}
where $E$ is the mass of the string (we neglect the quark
masses)
\begin{equation}
E\ \sim\ V(L)\ =\ TL\ ,
\end{equation}
while $\omega$ is the frequency of the angular rotation.
Expressing $\omega$ in terms of $j$ using $j\sim E\omega L^2$
we obtain
\begin{equation}
L^2\ \sim\ \frac jT\ .
\end{equation}

Using (4.17) again we get that the meson mass is
\begin{equation}
E^2\ \sim\ Tj\ .
\end{equation}
The mass square is proportional to $j$ at large $j$.

More precisely the spectrum of a free Nambu--Goto string goes
as
\begin{equation}
E^2\ =\ E^2_0+T(j+n)\ .
\end{equation}
Here $E^2_0$ is the intercept and $n$ labels the daughter
trajectories. At large $j$, $j\gg1$ our naive estimate (4.19)
gives the same result as the exact string spectrum (4.20).

 The
string tension $T$ is given by (4.5) to the leading order in
$\mu $. Expressed in terms of the photon mass (2.3) (or
Higgs mass, $m_\gamma\simeq m_H$) it reads
\begin{equation}
T\ =\ \frac{\pi m^2_\gamma}{n^2_eg^2}\ .
\end{equation}
We see that in terms of photon mass $T$ is large in the weak coupling
 $g^2\ll 1$. This is a typical result for solitonic objects
in the semiclassical approximation.

Now let us discuss the region of validity of (4.19) and (4.20).
The string theory result (4.20) assumes the approximation of long
and thin strings. The transverse size of the ANO vortex is given
by $1/m_\gamma$. So we need
\begin{equation}
L^2\ \gg\ \frac1{m^2_\gamma}\ .
\end{equation}
Substituting (4.18) and (4.21) here we get
\begin{equation}
j\ \gg\ \frac1{g^2}\ .
\end{equation}
We see that in fact Regge trajectories become linear only at
extremely large $j$.

The bound (4.23) is rather restrictive. If $j$ is not that large
the transverse size of the vortex becomes of  order of
its length and the string is not developed. Quark and anti-quark
are on the border between the stringy regime and the Coulomb
regime.
 At small $j$ $\tilde qq$-meson more looks like spherically
symmetric soliton  rather than a
string.

We can also see the breakdown of the string description from the
string representation for the ANO vortex. It is developed in
\cite{O,BS} and \cite{Y} for the cases of strings in the type II
superconductor, BPS-strings and strings in the type I
superconductor respectively. The common feature of these
representations is that the leading term of the world sheet
action is the Nambu--Goto term
\begin{equation}
S_{\rm string}=\ T\int d^2x\left\{\sqrt g+\ \mbox{ higher
derivatives }\right\},
\end{equation}
where $g_{ij}=\partial_ix_\mu\partial_jx_\mu$ is the induced metric
$(i,j=1,2)$. Higher derivative corrections in (4.24) contain
term important for the string quantization \cite{PS,ACPZ},
rigidity term \cite{P} etc. These terms contain  powers of
$\partial/m_\gamma$. For thin strings $m_\gamma\sim
m_H\to\infty$ and these corrections can be neglected in the action (4.24).
However, for the ANO vortex in the semiclassical regime
$\partial^2/m^2_\gamma\sim T/m^2_\gamma\sim1/g^2\gg1$ (see
(4.21)). Hence, higher derivative corrections blow up in (4.24)
and the string approximation is no longer acceptable. From the
string theory point of view this manifest itself as a "crumpled"
string surface \cite{P,D}.

We see that Regge trajectories are not linear in the wide region
of spins $j\la1/g^2$. This to be contrasted with perfect
linear behavior of Regge trajectories in Nature
starting from
small $j$ \footnote{ On the contrary, in QCD is  hard to
talk about linear trajectories at large $j$ because higher
resonances acquire large widths.} (for a recent account see
lecture \cite{Anis} given
at this School).

 It might seem funny that we are trying to
compare some properties of Seiberg--Witten theory with
experiment.   Still I believe that the linear behavior of Regge
trajectories is an important feature of 
confinement in QCD and we have to reproduce it in a theory
with QCD-like confinement.

The main property responsible for this "disadvantage" of the
confinement in the monopole point of the Seiberg--Witten theory
is the large value of the string tension (4.21). In the
conclusion of this talk I will speculate that this problem as
well as some others could be resolved if the string were
(almost) tensionless.

\subsection{Heavy quark-anti-quark states}

Another unpleasant consequence of the large value of the string
tension (4.21) is that hadrons built of quarks appear to be too
heavy.

To show this let me summarize qualitatively the low lying hadron
spectrum in our theory. First,  there are states with color
(magnetic) charge screened by the Higgs mechanism. They are
monopoles (described by operators
$\widetilde MM$) and the dual photon with its superpartners.
Their masses are of order of $m_\gamma$ $(m_H\approx m_\gamma$
at small $\mu $).

Second, there are hadrons built of quarks via the confinement
mechanism described above. As an example, consider
 $\tilde qq$-meson at $j\sim1$. Its mass is of order of
\begin{equation}
m_{\tilde qq}\ \sim\ 2m_q+\sqrt T\ ,
\end{equation}
where $m_q$ is the quark mass.

The problem is that the mass in (4.25) is too large (as compared with
$m_\gamma$) as I will show below. This means that we have
light monopole states and the photon (which we can
interpret  as glueballs), while states built of quarks are
heavy.  In contrast, in QCD we have light $\tilde qq$-states,
whereas the candidates for glueballs are much heavier.

First, let us see how small $m_q$ in (4.25) can be. So far
we  discussed the pure gauge theory or the theory with very
heavy quarks which we used as a probe for the confinement. Now
let us introduce one dynamical flavor  of the
fundamental matter hypermultiplet
\cite{SW2}(we call it quark).

The quark mass on the Coulomb branch of ${\cal N}=2$ theory is given by
\begin{equation}
m_q\ =\ m+\frac a{\sqrt2}\ ,
\end{equation}
where $m$ is the quark mass parameter in the microscopic theory.
Quarks become light near the charge singular point on the
Coulomb branch, $a=-\sqrt2\,m$. Hence, in order to have light
quarks we have to go near the charge singularity.

On the other hand, once we switch on $\mu $ the Coulomb
branch shrinks to three singular points: monopole, dyon and
charge ones. As we discuss before, in order to have monopole
condensation and quark confinement we have to go to the monopole
point.

Now to make quarks light near the monopole point we  choose  $m$
to ensure that the charge singularity goes close on the Coulomb
branch to the monopole one. The values of $m$ corresponding to
the colliding of these singularities are called Argyres--Douglas
(AD) points \cite{AD}. In $SU(2)$ theory these points were
studied in \cite{APSW}. On the Coulomb branch of ${\cal N}=2$ theory
these points flow in the infrared to a non-trivial conformal field
theories.

The value of $m$ at which the monopole singularity collides with the
charge one is \cite{APSW} (there are three such points; we choose
one of them corresponding to real $m_{AD}$)
\begin{equation}
m_{AD}\ =\ \frac3{2^{4/3}}\ \Lambda_1\ ,
\end{equation}
where $\Lambda_1$ is the scale of the theory with one flavor,
$\Lambda^4=m\Lambda^3_1$ at large $m$ .

The problem, however, is that we cannot go directly to the
AD-point (4.27). The point is that the monopole condensate
vanishes at the AD-point \cite{GVY}. This means that the
AD-point is the point of the quark deconfinement \cite{GVY}.

Now the question is whether we can go close to the AD-point to
make quarks light enough without destroying the monopole
condensation and quark confinement. To answer this question let
us develop a perturbation theory around AD-point.

In general, the monopole condensate in the theory with one
flavor is \cite{GVY}
\begin{equation}
\langle\widetilde MM\rangle\ =\
2i\mu \left(u^2_m-2m\Lambda^3_1\right)^{1/4}\ ,
\end{equation}
where $u_m$ is the position of the monopole singularity in the
$u$-plane given by the Seiberg--Witten curve \cite{SW2}. In
particular, in the AD-point $u^{AD}_m=\frac43m^2_{AD}$. This
value makes  the monopole condensate in (4.28) vanish.

Now let us take $m$ close to its AD-value
\begin{equation}
m\ =\ m_{AD}(1+\varepsilon)\ ,
\end{equation}
where $\varepsilon\ll1$. Then extracting $u_m$ from the
Seiberg--Witten curve near AD point \cite{GVY} we get
\begin{equation}
\langle\widetilde MM\rangle\ \sim\ \mu \Lambda_1
\varepsilon^{1/4}\ .
\end{equation}

The value of the monopole condensate (4.30) set the scale of our
effective Abelian Higgs model. In particular, the photon mass
and ANO string tension is given by Eqs. (2.3) and (2.8), where
the Higgs VEV  $v^2$ is identified with the monopole condensate (4.30).

Now let us see if we can make quark mass in (4.25) to be small
as compared to the string scale $T^{1/2}\sim\langle\widetilde
MM\rangle^{1/2}$. According to ref.\cite{APSW} the anomalous
dimension of $m_q$ in (4.26) is one, while the anomalous
dimension of $m-m_{AD}$ is $4/5$. Using this we conclude that
\begin{equation}
m_q\ \sim\ \Lambda_1\varepsilon^{5/4}\ .
\end{equation}
From (4.30) and (4.31) we see that we  can always make quarks
lighter than the string scale $T^{1/2}$ if we take $\mu $
not too small, $\mu \gg\Lambda_1\varepsilon^{9/4}$. Of course, we
still keep $\mu \ll\Lambda_1$ to ensure the validity of our
dual QED description.

Unfortunately, this does not solve the problem of heavy
$\tilde qq$-states .
 As we already mentioned, the string scale is much
larger or at least of the same order as  the photon mass,  (see (4.21)).
Therefore, the mass of $\tilde qq$-meson in (4.25) is much larger
 or of the same order
as the photon and  monopole  masses in
contrast with our expectations about the theory with QCD-like
confinement.

\section{Confinement on  Higgs branches}
\setcounter{equation}{0}

In this section I will consider another confinement scenario in
the Seiberg--Witten theory: confinement on the Higgs branch in
the theory with matter. I will  talk about $SU(2)$ theory
with $N_f=2$ flavors of fundamental matter \cite{SW2}. The
$N_f=2$ case is the simplest case of the theory with Higgs
branches  (for $N_f=1$ we do not have Higgs branches).

\subsection{Review of Higgs branches in SU(2) theory}

Let us introduce $N_f=2$ fundamental matter hypermultiplets in
the ${\cal N}=2$ SU(2) gauge theory. In terms of ${\cal N}=1$ superfields
matter dependent part of the microscopic action looks like
\begin{eqnarray}
&& S_{\rm matter}=\int d^4xd^2\theta d^2\bar\theta\left[\bar Q_A
e^VQ^A+\bar{\widetilde  Q}^Ae^V\widetilde Q_A\right]\ + \nonumber\\
+&& i\int d^4xd^2\theta\left[\sqrt2\widetilde Q_A\frac{\tau^a}2\
Q^A\Phi^a+m\widetilde Q_AQ^A\right]+\mbox{ c.c.}
\end{eqnarray}
Here $Q^{kA},\widetilde Q_{Ak}$ are matter chiral fields, $k=1,2$ and
$A=1,\ldots,N_F$, while $V$ is the vector superfield.
Thus we have 16 real matter degrees of freedom for $N_f=2$.

Consider first the limit of large $m$. In this limit the  three
singularities on the Coulomb branch are easy to understand. Two
of them correspond to monopole and dyon singularities of the
pure gauge theory. Their positions on the Coulomb branch are
given by \cite{SW2}
\begin{equation}
u_{m,d}\ =\ \pm\ 2m\Lambda_2-\frac12\Lambda^2_2\ ,
\end{equation}
where $u=\frac12\langle\varphi^{a^2}\rangle$ and $\Lambda_2$ is
the scale of the theory with $N_f=2$.
In the large $m$ limit $u_{m,d}$ are
approximately given by their values in the pure gauge theory
$u_{m,d}\simeq\pm2m\Lambda_2=\pm2\Lambda^2$, where $\Lambda$
is the scale of $N_f=0$ theory.

The third singularity corresponds to the point where charge
becomes massless. Let us decompose matter fields as
\begin{equation}
Q^{kA}\ =\ \left({1\atop 0}\right)^k Q^A_++\left({0\atop 1}
\right)^kQ^A_-\ .
\end{equation}
From the superpotential in (5.1) we see that the $Q_+$ becomes
massless at
\begin{equation}
a\ =\ -\ \sqrt2\ m\ .
\end{equation}
The singular point $a=+\sqrt2\,m$ is gauge equivalent to the one
in (5.4). In terms of variable $u$ (5.4) reads
\begin{equation}
u_c\ =\ m^2+\frac12\ \Lambda^2\ .
\end{equation}
Strictly speaking, we have $2+N_f=4$ singularities on the
Coulomb branch. However two of them  coincides for the case of
two flavors of matter with the same mass.

The effective theory on the Coulomb branch near charge
singularity (5.4) is given by ${\cal N}=2$ QED with light matter fields
$Q^A_+$, $\widetilde Q_{+A}$ (8 real degrees of freedom) as well as the
photon multiplet.

The charge singularity (5.4),(5.5) is the root of the Higgs
branch \cite{SW2}. To find this branch let us write down $D$-term
and $F$-term conditions which follow from (5.1). $D$-term
conditions are
\begin{equation}
Q^{kA}\bar Q_{A\ell}+\bar{\widetilde Q}^{kA}\widetilde Q_{A\ell}\ =\ 0\ ,
\end{equation}
while $F$-term conditions give (5.4) as well as
\begin{equation}
Q^{kA}\widetilde Q_{A\ell}\ =\ 0\ .
\end{equation}
Eqs. (5.6),(5.7) have nontrivial solutions for $N_f\ge2$. These
solutions determines VEV's for scalar components  $q^{kA}$, $\widetilde
q_{Ak}$ of fields $Q^{kA}, \widetilde Q_{Ak}$. Dropping heavy
components $q_-$ according to decomposition (5.3) and
introducing the SU$_R(2)$ doublet $q^{fA}$ as
\begin{eqnarray}
q^{1A}=\ q^A_+\ , && q^{2A}=\ \bar{\widetilde q}^A_+\ , \nonumber\\
\bar q_{A1}=\ \bar q^+_A\ , && \bar q_{A2}=\ -\widetilde q^+_A\ ,
\end{eqnarray}
we can rewrite three real conditions in (5.6),(5.7) as
\begin{equation}
\bar q_{Ap}(\tau^a)^p_f\ q^{fA}\ =\ 0, \quad a=1,2,3.
\end{equation}
Eq.(5.9) together with the condition (5.4) determines the Higgs
branch (manifold with $\langle q\rangle\neq0$) which touches the
Coulomb branch at the point (5.4).

The low energy theory for boson fields near the root of the Higgs
branch looks like
\begin{equation}
S^{\rm root}_{\rm boson}=\int d^4x\left\{\frac1{4g^2}
F^2_{\mu\nu}+\bar\nabla_\mu\bar q_{Af}\nabla_\mu q^{fA}+
\frac{g^2}8[\mbox{ Tr }\bar q\tau^aq)]^2\right\},
\end{equation}
where trace is calculated over flavor and SU$_R(2)$ indices.
Here $\nabla_\mu=\partial_\mu-in_eA_\mu$, $\bar\nabla_\mu=
\partial_\mu+in_eA_\mu$, the electric charge $n_e=1/2$ for
fundamental matter fields.

This is an Abelian Higgs model with last interaction term coming
from the elimination of $D$ and $F$ terms. The QED coupling
constant $g^2$ is small near the root of the Higgs branch.
 We include 8 real matter degrees of
freedom $q^{fA}$ in the theory (5.10) according to the
identification (5.8). The rest of matter fields $q^A_-$, $\widetilde
q\,^-_A$ (another 8 real degrees of freedom) acquire a large mass
$2m$ and can be dropped out. The effective theory (5.10) is
correct on the Coulomb branch near the root of the Higgs branch
(5.4) or on the Higgs branch not far away from the origin
$\langle q\rangle=0$.

It is clear that the last term in (5.10) is zero on the fields
$q$ which satisfy constraint (5.9). This means that moduli
fields which develop VEV's on the Higgs branch are massless, as
it should be. The other fields acquire mass of order
$\langle\bar qq\rangle^{1/2}$. It turns out that there are four
real moduli fields $q$ (out of 8) which satisfy the constraint
(5.9) \cite{SW2}. They correspond to the lowest components of  one
hypermultiplet.

We can parameterize them as
\begin{equation}
q^{f\dot A}(x)\ =\ \frac1{\sqrt2}\ \sigma^{f\dot A}_\alpha
\phi_\alpha(x) e^{i\alpha(x)}\ .
\end{equation}
Here $\phi_\alpha(x), \alpha=1\ldots4$ are four real moduli
fields. It is clear that fields (5.11) solve (5.9). The common
phase $\alpha(x)$ in (5.11) is the U(1) gauge phase. Once
$\langle\phi_\alpha\rangle=v_\alpha\neq0$ on the Higgs branch the U(1)
group is broken and $\alpha(x)$ is eaten by the Higgs  mechanism.
Say, in the unitary gauge $\alpha(x)=0$. In the next subsection
we consider vortex solution for the model (5.10). Then
$\alpha(x)$ is determined by the behavior of the gauge field at
the infinity. Substituting (5.11) into (5.10) we get the bosonic
part of the effective theory
for the massless moduli fields  on the Higgs branch near the origin
\begin{equation}
S^{\rm Higgs}_{\rm boson}\ =\ \int d^4x\left\{\frac1{4g^2}\
F^2_{\mu\nu}+\bar\nabla_\mu\bar q_\alpha\nabla_\mu
q_\alpha\right\}\ ,
\end{equation}
where
\begin{equation}
q_\alpha(x)\ =\ \phi_\alpha(x)\ e^{i\alpha(x)}\ .
\end{equation}

Once $v_\alpha\neq0$ we expect monopoles (they are heavy at
$m\gg\Lambda$) to confine via formation of vortices which carry
the magnetic flux. The peculiar feature of the theory (5.12) is
the absence of the Higgs potential. Therefore, the Higgs phase of
of the theory in
(5.12) is the limiting case of type I superconductor with the
vanishing Higgs mass. In the next subsection I will consider the
peculiar features of ANO vortices in the model (5.12).

If we increase $v^2_\alpha$ taking $v^2_\alpha\ga\Lambda^2$ we
can integrate out massive photon. Then the effective theory is a
$\sigma$-model for massless fields $q_\alpha$ which belong to
4-dimensional Hyper--Kahler manifold, $R^4/Z_2$. The metric of
this $\sigma$-model is flat \cite{SW2,APS}, there are, however,
higher derivative corrections induced by instantons \cite{Y2}. 
Here we consider region of Higgs branch with
$v^2_\alpha\ll\Lambda_2$. This determines the scale of the
effective Abelian Higgs model (5.12). $W$-bosons and other
particles which reflect the non-\-Abelian structure of the
underlying microscopic theory are heavy with masses
$\ga\Lambda_2$ and can be ignored.

To conclude this subsection let us briefly review what happens if
we reduce the mass parameter $m$. At $m=\pm\Lambda$ the charge
singularity (root of the Higgs branch) collides with the monopole
(dyon) singularity, see Eqs.(5.2),(5.5). These are
Argyres--Douglas points \cite{AD,APSW}. At these points mutually
non-\-local degrees of freedom (say, charges and monopoles)
becomes massless simultaneously. These points are very
interesting from the point of view of the monopole confinement on
the Higgs branch. Mono\-pol\-es become
dynamical as we approach Argyres--Douglas point, $m\to\Lambda_2$.

After the collision quantum numbers of particles at
singularities change because of monodromies \cite{SW2}. If we
denote quantum numbers as $(n_m,n_e)_B$, where $n_m$ and $n_e$
are magnetic and electric charges of the state, while $B$ is its
baryon number then at $m>\Lambda_2$ we have charge, monopole and
dyon singularities with quantum numbers
\begin{equation}
(0,\ 1/2)^{*2}_1\ , \quad (1\ 0)_0\ , \quad (1,\ 1)_0\ .
\end{equation}
The superscript for the charge means that we have two flavors
of charges. After charge singularity collides with monopole one
(at $m<\Lambda_2)$ the quantum numbers of particles at
singularities become \cite{BF}
\begin{equation}
(1,\ 0)^{*2}_0\ , \quad (1,\ 1/2)_1\ , \quad (1,\ 1/2)_{-1}\ .
\end{equation}
Now monopole $(1,0)_0$ condense on the Higgs branch which
emerges from the point (5.5), while dyons $(1,1/2)_1$ and
$(1,1/2)_{-1}$ confine because they carry electric
charge. At
zero mass, $m=0$ two dyon singularities in (5.15) coincide (see
(5.2)) and the second Higgs branch appears at the point
$u=-1/2\,\Lambda^2_2$. This restores the global symmetry from
SU$(N_f=2)$ in the massive theory to SO$(2N_f=4)$ at $m=0\ $
\cite{SW2}.

\subsection{ANO string on the Higgs branch}

Now let us focus on confinement of monopoles on the Higgs branch
and consider ANO vortices in the model (5.12) 
\footnote{ Let us take
$m>\Lambda_2$ to avoid confusion with notation of dyon states.}.
This is done in \cite{Y}.

Without loss of generality we take VEV's of $q_\alpha$
$v_\alpha=(v,0,0,0)$. Moreover, we drop fields $q_2,q_3$ and
$q_4$ from (5.12) because they are irrelevant for the purpose of
finding classical vortex solutions. Thus, we arrive at the
Abelian Higgs model (2.1) with $\lambda=0$ and identification
$q_1=\varphi$.

Following \cite{Y} consider first the model (2.1) with small
$\lambda$, so that $m_H\ll m_\gamma$ (see (2.3), (2.4) for
$n_e=1/2$). Then  we take the limit $m_H\to0$.

To the leading order in $\log m_\gamma/m_H$ the vortex
solution has the following structure \cite{Y}. The
electromagnetic field is confined in a core with the radius
\begin{equation}
R^2_g\ \sim\ \frac1{m^2_\gamma}\ln^2\frac{m_\gamma}{m_H}\ .
\end{equation}
The scalar field is close to zero inside the core. Instead,
outside the core, the electromagnetic field is vanishingly
small. At intermediate distances
\begin{equation}
R_g\ \ll\ r\ \ll\ \frac1{m_H}
\end{equation}
($r$ is the distance from the center of vortex in (1,2) plane)
the scalar field satisfy the free equation of motion. Its
solution reads \cite{Y}
\begin{equation}
\varphi(r)\ =\ v\left(1-\frac{\ln1/r\,m_H}{\ln
1/R_g\,m_H}\right) .
\end{equation}
At large distances $r\gg1/m_H$~~ $\varphi$ approaches its
VEV as $\varphi-v \sim\exp(-m_Hr)$.

The main contribution to the string tension comes from the
logarithmically large region (5.17), where scalar field is given
by (5.18). The result for the string tension is \cite{Y} (as we
already mentioned, see (2.9))
\begin{equation}
T\ =\ \frac{2\pi v^2}{\ln m_\gamma/m_H}\ .
\end{equation}
It comes from the kinetic energy of the scalar field in (2.1)
("surface" energy).

The results in (5.16),(5.19) mean that if we naively take the
limit $m_H\to0$ the string becomes infinitely thick and its
tension goes to zero \cite{Y}. This means that there are no
strings in the limit $m_H=0$. The ANO vortex becomes a vacuum
state with "twisted" boundary conditions which ensure the
magnetic flux $2\pi/n_e$.~~\footnote{The author is indebted to
A.~Vainshtein for this interpretation.} The absence of ANO
strings in theories with flat Higgs potential was noticed in
\cite{R,ARH}. ANO strings on the Higgs branch was also discussed
in \cite{GMV} from the brane point of view. However it was not
noticed in \cite{GMV} that the vortex becomes infinitely "thick"
and its tension goes to zero.

One might think that the 
 absence of ANO strings means that there is no
  confinement on Higgs branches. As
we will see now this is not  the case \cite{Y}.

So far we have considered infinitely long ANO strings. However
the setup for the confinement problem is slightly different
\cite{Y}. We have to consider monopole--anti-monopole pair at
large but finite separation $L$. Our aim is to take the limit
$m_H\to0$. To do so let us consider ANO string of the finite
length $L$ within the region
\begin{equation}
\frac1{m_\gamma}\ \ll\ L\ \ll\ \frac1{m_H}\ .
\end{equation}
Then it turns out that $1/L$ plays the role of the $IR$-cutoff
 in Eqs. (5.16) and (5.19) instead of $m_H$ \cite{Y}.

The reason for this is easy to understand. In fact, the reason
for the absence of the vortex solution is quite clear from
(5.18). If we put $m_H$ to zero the scalar field $\varphi$
cannot reach its VEV at infinity because of its logarithmic
behavior. This was noticed in \cite{R}.

Now consider vortex of finite length $L$ and look at the
behavior of $\varphi$ at large distances $r$, $L\ll r\ll1/m_H$.
In this region the problem becomes three dimensional. The
solution of the free equation of motion in three dimensions
reads
\begin{equation}
\varphi-v\ \sim\ \frac1{|x|}\ ,
\end{equation}
where $|x|$ is three dimensional distance from the vortex. The
solution (5.21) perfectly goes to its VEV $v$ at infinity. To
put it in the other way at distances $|x|\gg L$ from the vortex
the two dimensional problem transforms into the three dimensional one.
Namely, the scalar field has logarithmic behavior at $r$ within the 
bounds  $R_g\ll r\ll L$, whereas at $|x|\gg L$ it acquires $1/|x|$
behavior given by (5.21).
We see that $L$ really plays the role of the $IR$-cutoff for the
logarithmic behavior of $\varphi$. Now we can take the limit $m_H\to 0$.

The result for the electromagnetic core of the vortex becomes
\cite{Y}
\begin{equation}
R^2_g\ \sim\ \frac1{m^2_\gamma}\ln^2m_\gamma L\ ,
\end{equation}
while its string tension is given by \cite{Y}
\begin{equation}
T\ =\ \frac{2\pi v^2}{\ln m_\gamma L}\ .
\end{equation}

We see that the ANO string becomes "thick" but still its
transverse size $R_g$ is much less than its length $L$, $R_g\ll
L$. As a result the potential between heavy well separated
monopole and anti-monopole is still  confining but is
no longer linear in $L$. It behaves
as \cite{Y}
\begin{equation}
V(L)\ =\ 2\pi v^2\, \frac L{\ln m_\gamma L}\ .
\end{equation}
 As soon as the
potential $V(L)$ is an order parameter which distinguishes 
different phases of a theory
(see, for example, review
\cite{IS}) we conclude that we have a new confining phase on the
Higgs branch of the Seiberg--Witten theory. It is clear that
this phase can arise only in supersymmetric theories
because we do not have Higgs branches without supersymmetry.

Unfortunately, the confinement on Higgs branches cannot play a
role of a model for QCD-like confinement. In particular, the
confining potential (5.24) gives rise to the following behavior
of Regge trajectories at ultra-large $j$, $j\gg1/g^2$ (see
(4.18))
\begin{equation}
E^2\ \sim\ v^2\frac j{\ln(g^2j)}\ .
\end{equation}
We see that Regge trajectories never becomes linear. At
$j\la1/g^2$ they are non-linear due to the reason we have
discussed in subsection 4.3. At $j\gg1/g^2$ they are still
non-linear because of the logarithmic factor in (5.25).

It is worth note also that because of type I superconductivity 
on the Higgs branch we
have a tower of ``exotic'' hadron states corresponding to higher
winding numbers of strings. 

\section{Outlook: ~~ tensionless strings}
\setcounter{equation}{0}

We have discussed two confinement scenarios in the
Seiberg--Witten theory: the confinement of quarks in the monopole
point upon breaking ${\cal N}=2$ down to ${\cal N}=1$ and the confinement of
monopoles on the Higgs branch in ${\cal N}=2$ theory. The latter one
does not looks like confinement in QCD, while the former one is
more promising.

Still as we have discussed in Section 4 it has several unwanted
features. Basically these unwanted features fall into two
categories. First, is the presence  of extra states in the
hadron spectrum. It is a reflection of $U(1)$ nature of the
confinement in Seiberg--Witten theory. The second is associated
with the large value of the string tension (4.21) as compared
with the squire of the photon or Higgs mass.

One can believe that the first problem disappears if we increase
$\mu $ and go to the limit of ${\cal N}=1$ QCD. Now I would like to
speculate that the second problem also might disappear in the
same limit.

One might think that once we increase $\mu $ the dual QED
coupling $g^2$ increases and becomes large, $g^2\gg 1$.
 If this happens the
string tension becomes small, see (4.21). As a result Regge
trajectories becomes linear already at $j\sim1$. Moreover,
hadrons built of quarks becomes light and can be visible in the low
energy spectrum. Instead monopoles and photon (to be interpreted
as glueballs) become relatively heavy. Of course, at strong coupling
we loose control over the theory and cannot use eq. (4.21) relating
the string tension to the photon mass. The string is not a BPS
state and its tension is not protected from corrections. Moreover,
we do not have exact formula for the photon mass in the strong coupling as
well. Still we can speculate that there is a region in the parameter
space where the string tension becomes much less then the 
squire of the photon mass.

In fact, the conjecture that ANO strings might become
tensionless in the strong coupling limit of QED was suggested in
Ref.\cite{HK} in order to find the field theory explanation of
tensionless $M$-theory strings which arise when 5-branes
approach each other.

Note, that we have to find the regime such that $T$ goes to
zero, while $m_\gamma$ and $m_H$ (which control the transverse
size of ANO string) stay finite.

For example, the string on a Higgs branch discussed in the
previous section does not do the job. Its tension goes to zero
but its transverse size $R_g$ becomes infinite (see (5.16)).
Thus this string disappears. Another example is the string in the
monopole point at the $AD$-value of the quark mass (see
subsection 4.4). As the monopole condensate vanishes the string
tension goes to zero, however the string transverse size
might  become
infinite. It is not clear if this string  can serve as a QCD string.

To have a QCD-like confinement we need a  regime in
which string remains a string (its size does not grow) while its
tension becomes small.

\subsection*{Acknowledgments}

The author is grateful to the organizers of XXXIV PNPI Winter school for
the opportunity to present this lecture. Also the author  would like to
thank A.~Hanany, M.~Shifman, M.~Strassler, and, in particularly,
A.~Vainshtein for useful discussions.
The author is also grateful to the Theoretical Physics Institute at the 
University of Minnesota for support.
 This work is also supported by
Russian Foundation for Basic Research under grant
No.99-02-16576.


\begin{thebibliography}{99}
\bibitem{1}  C.~Montonen and D.~Olive, Phys.Lett.B {\bf72}
(1977) 117;\\
P.~Goddard, J.~Nuyts and D.Olive, Nucl.Phys.B {\bf125} (1977) 1.

\bibitem{SW1}N.~Seiberg and E.~Witten, hep-th/9407087;
Nucl.Phys.B {\bf426} (1994) 19.

\bibitem{SW2}N. Seiberg and E. Witten, hep-th/9408099;
Nucl.Phys.B {\bf431} (1994) 484.

\bibitem{MH}S. Mandelstam, Phys.Rep. {\bf23C} (1976) 145;\\
G.'t Hooft, in Proceed.of the Europ.Phys.Soc. 1975, ed.by
A.Zichichi (Editrice Compositori, Bologna, 1976), p.1225.

\bibitem{ANO}A.A. Abrikosov, Sov.Phys.JETP, {\bf32} (1957)
1442;\\
H.B. Nielsen and P. Olesen, Nucl.Phys. {\bf B61} (1973) 45.

\bibitem{Y} A. Yung, hep-th/9906243; Nucl.Phys. B{\bf562} (1999)
191.

\bibitem{B}E.B. Bogomolny, Sov.J.Nucl.Phys. {\bf24} (1976) 449.

\bibitem{HS}Z. Hlousek and D. Spector, Nucl.Phys.B {\bf370}
(1992) 143;\\
J. Edelstein, C.Nunez, and F.~Schaposnik,, hep-th/9311055;
Phys.Lett. {\bf329B} (1994) 39.

\bibitem{BV}E.B. Bogomolny and A.I. Vainshtein, Sov.J.Yad.Fiz.
{\bf23} (1976) 1111.

\bibitem{HSZ} A. Hanany, M. Strassler, and A.~Zaffaroni,
hep-th/9707244; Nucl.Phys.B {\bf513} (1998) 87.

\bibitem{S} M. Strassler, hep-lat/9803009; Prog. Theor. Phys.
Suppl. {\bf131} (1998) 439.

\bibitem{VY} A. Vainshtein and A. Yung, {\em"Type I
superconductivity at the monopole point in the Seiberg--Witten
theory"}, to appear.

\bibitem{FI} P. Fayet and J. Iliopoulos, Phys. Lett. {\bf51}B
(1974) 461.

\bibitem{DDT} S.C. Davis, A. Davis, and M. Trodden,
hep-th/9702360; Phys. Lett. B{\bf405} (1997) 257.

\bibitem{GS} A. Gorsky and M. Shifman, 
 hep-th/9909015; Phys. Rev. D {\bf61} (2000) 08 5001.

\bibitem{Sp}W. Fuertes and J. Guilarte,  hep-th/9807218; Phys.Lett.B
{\bf437} (1998) 82.

\bibitem{W} E. Witten, hep-th/9706109; Nucl.Phys. B{\bf507}
(1997) 658.

\bibitem{DS} M.R. Douglas and S.H. Shenker, hep-th/9503163;
Nucl.Phys. B{\bf447} (1995) 271.

\bibitem{Sp2}J.D. Edelstein, W.G. Fuertes, J. Mas and
J.M.~Guilarte, {\em ``Phases of dual superconductivity and
confinement in softly broken {\cal N}=2 supersymmetric Yang--Mills
theories''}, hep-th/0001184.


\bibitem{O} P. Orland, hep-th/9404140; Nucl.Phys. B{\bf428}
(1994) 221;\\ M. Sato and S. Yahikozawa, hep-th/9406208;
Nucl.Phys.B {\bf436} (1995) 100.


\bibitem{BS}M. Baker and R. Steinke, hep-ph/9905375;
Phys. Lett. B {\bf474} (2000) 67


\bibitem{PS}J. Polchinski and A. Strominger, Phys.Rev.Lett.
{\bf67} (1991) 1681.

\bibitem{ACPZ}E. Akhmedov, M. Chernodub, M.~Polikarpov, and
M.~Zubkov, Phys.Rev.D {\bf53} (1996) 2087.

\bibitem{P}A. Polyakov, Nucl.Phys.B {\bf268} (1986) 406.

\bibitem{D}G. David, Phys.Rep. {\bf184} (1989) 221.

\bibitem{Anis} A.V. Anisovich, V.V. Anisovich
 and A.V.~Sarantsev, {\em ``Systematics of $q\bar
q$-states in $(h,M^2$) and $(J,M^2)$ planes"},
hep-ph/0003113; Proceedings of
the XXXIV PNPI Winter School,Repino, St. Petersburg, 2000.

\bibitem{AD}P.C. Argyres and M.R. Douglas, hep-th/9505062;
Nucl.Phys.B {\bf448} (1995) 93.

\bibitem{APSW}P.C. Argyres, M.R. Plesser, N.~Seiberg, and
E.~Witten, hep-th/9511154; Nucl.Phys.B {\bf461} (1996) 71.

\bibitem{GVY} A. Gorsky, A. Vainshtein and A. Yung, {\em
``Deconfinement at the Argyres-Douglas point in SU(2) gauge theory
with broken {\cal N}=2 supersymmetry''}, hep-th/0004087.

\bibitem{APS} P. Argyres, M. Plesser and N. Seiberg, hep-th/9603042;
Nucl. Phys. B {\bf471} (1996) 159.

\bibitem{Y2} A. Yung, hep-th/9705181; Nucl. Phys. B {\bf512} (1998) 79

\bibitem{BF} A. Bilal and F. Ferrari, hep-th/9706145;
Nucl.Phys.B {\bf516} (1998) 175.

\bibitem{R}A. Penin, V. Rubakov, P.~Tinyakov and S.~Troitsky,
hep-th/9609257; Phys.Lett B {\bf389} (1996) 13.

\bibitem{ARH} A. Achucarro, M. de Roo and L.~Huiszoon,
Phys.Lett. B{\bf424} (1998) 288.

\bibitem{GMV} B.R. Greene, D.R. Morrison and C.~Vafa,
hep-th/9608039; Nucl.Phys.B {\bf481} (1996) 513.

\bibitem{IS} K. Intrilligator and N. Seiberg, hep-th/9509066;
Nucl.Phys. B (Proc. Suppl.) {\bf45}BC (1996).

\bibitem{HK}A. Hanany and I. Klebanov, hep-th/9606136;
Nucl.Phys.B {\bf482} (1996) 105.




\end{thebibliography}
\end{document}